\documentclass[aps,nofootinbib,longbibliography,prd,onecolumn]{revtex4}
\usepackage[a4paper,top=2cm,bottom=2cm,left=2cm,right=2cm]{geometry}
\usepackage[babel]{csquotes}
\usepackage{graphicx}
\usepackage{amsmath,amssymb,amsfonts}
\usepackage[applemac]{inputenc}
\usepackage{mathtools,mathrsfs}
\usepackage{enumerate}
\usepackage{dsfont}
\usepackage{scrextend}
\usepackage{float}
\usepackage{MyCommands}
\usepackage[colorlinks,citecolor=blue,linkcolor=blue,urlcolor=blue]{hyperref}

\begin{document}

\title{A Noether Theorem for discrete Covariant Mechanics}
\author{Fabio D'Ambrosio}\email{fabio.dambrosio@gmx.ch}

\affiliation{\small
\mbox{Centre de Physique Th\'eorique, Aix--Marseille Universit\'e, Marseille, France.} }
\date{\small\today}
\begin{abstract}
\noindent
	Noether's theorem is an elegant and powerful tool of classical mechanics, but it is of little to no consequence in discrete theories. Here we define and explore a discrete approach to covariant mechanics and show that within this framework a discrete version of Noether's theorem, completely analogous to the well-known continuum version with all its ramifications, remains valid. We also discuss why more traditional approaches to discretized mechanics violate certain conservation laws by construction.
\end{abstract}

\maketitle

\section{Introduction}
Discretization plays a major role in many areas of physics and engineering. On one hand, it constitutes the basis for numerical algorithms aimed at solving complex problems in mechanics, fluid dynamics, electrodynamics and general relativity. On the other hand, discretization is also important in certain branches of theoretical physics. Richard Feynman \cite{Feynman1948}, for example, relied on discretization techniques to define and construct his path integral which in the course of time lead to great advances in the understanding of QED and QCD \cite{Wilson}. Lattice QCD can even be understood as a \textit{definition} of the theory of strong interactions, rather than a mere discretization of a continuum theory \cite{QCD_Book}. Similarly, the spin foam approach to quantum gravity uses simplicial discretizations as a building block to define a theory of quantum gravity \cite{EPRL1,EPRL2,FK,RovelliNewBook}. More recently, it has even been argued that discrete theories provide a minimal yet complete description of elementary physics \cite{Maudlin,Rovelli2019}.

Despite their many uses and applications, discretization techniques have their limitations. One of them is that quantities which are conserved in the continuum theory may no longer be conserved in the discretized theory. Occasionally, this failure of the discrete theory is falsely attributed to a breaking of continuous symmetry induced by the process of discretization. A lattice discretization of three dimensional Euclidean space, for example, manifestly breaks $SO(3)$ symmetry. However, despite this loss of symmetry, angular momentum is conserved, as was first shown by Baez and Gilliam in \cite{Baez1994}. In fact, Baez and Gilliam, and later Bahr, Gambini, and Pullin \cite{Bahr2011}, succeeded in proving a special case of Noether's theorem for discrete mechanical systems. Their theorem is applicable to coordinate transformations which act on the configuration space, while leaving the time parameter unaffected. Consequently, their theorem links conservation of momentum and angular momentum to the corresponding spatial symmetries in the discrete theory, but it makes no statements about systems with time translation symmetry or about systems subjected to Galilean  or Lorentzian boosts. 

It was later observed by Rovelli \cite{Rovelli2011} that the discretization of the covariant action for an harmonic oscillator leads to conservation of energy in the discrete theory. In this paper, we will show that this observation can be generalized to arbitrary (relativistic) mechanical systems, when expressed in a covariant language. We will formulate and prove a discrete version of Noether's theorem and thereby generalize the results of \cite{Baez1994,Bahr2011,Rovelli2011}.

In section~\ref{sec:CovMech} we review the covariant formulation of classical mechanics and we compare it to the more familiar Lagrangian formulation known from basic physics courses. In section~\ref{sec:Discrete} we then proceed to define and explore a discrete version of covariant mechanics. The main result of this section is the formulation and proof of a discrete version of Noether's theorem, followed by an illustrating example.
We then conclude the paper in section~\ref{sec:Conclusion} with a few comments and outlooks.

\section{Covariant Mechanics}\label{sec:CovMech}
Mechanics is traditionally understood as the study of the dynamics of one or several bodies who's configuration coordinates evolve in time $t$. The coordinate functions $q(t)$ are thereby regarded as dynamical variables while $t$ is a mere parameter used to describe evolution. However, an alternative description of mechanics is available, in which time is promoted to a dynamical variable \cite{DiracBook}. Evolution is now described by an arbitrary, unphysical parameter $s$. Position coordinates $q(s)$ and time coordinates $t(s)$ are treated on the same footing and, as we will see shortly, mechanics becomes a gauge theory. If this covariant formulation of mechanics seems to be somewhat artificial, it should be kept in mind that it arises naturally in the description of the relativistic point particle or that Maupertuis's principle is a covariant formulation of Newtonian mechanics in disguise (see Appendix~\ref{App_A}). 

It may also seem that the covariant formulation is more complicated than the more traditional formalisms of classical mechanics. However, it is the richer structure of the covariant theory which will allow us to formulate a discrete version of Noether's theorem. 

\subsection{Reparametrization invariant Action and Equations of Motion}
To describe a mechanical system in the more traditional Lagrangian formalism, one  starts by introducing a configuration space $\mathcal C$ and a Lagrangian $L=L(q(t),\dot q(t), t)$ defined on the tangent bundle $T\mathcal C$, where $q(t)$ are the spatial variables, $\dot q(t) := \frac{\dd q(t)}{\dd t}$, and $t$ is the time parameter.

In order to promote $t$ from a mere parameter to a genuine dynamical variable, we introduce the extended configuration space $\mathcal C_\textsf{ext}:= \mathcal C\times\mathbb R$, assumed to be an $n+1$ dimensional manifold coordinatized by $(q,t)$. By defining a space of trajectories, $\mathcal P:=\{(q,t):[s_i,s_f]\to\mathcal C_\textsf{ext} \, \vline\, (q,t)\in C^2(\mathbb R), \, \frac{\dd t}{\dd s}\neq 0\}$, we can proceed and introduce the reparametrization invariant action functional $S:\mathcal P\to\mathbb R$ defined by
\begin{equation}\label{eq:ActionFunctional}
	S[q, t] := \int_{s_i}^{s_f} L(q(s), \dot q(s)/\dot t(s), t(s)) \, \dot t(s)\, \dd s,
\end{equation}
where $s$ parametrizes the trajectory $(q,t)\in\mathcal P$ and the dot represents a total derivative with respect to $s$. Taking the variation of this action with respect to $q$ and $t$ results in the following two equations\footnote{A comment on notation: Expressions like $\PD{L}{q}$ are to be understood as $\nabla_q L$, since $q$ is a $n$-dimensional vector. The dot in $a\cdot b$ indicates a Euclidean scalar product between the vectors $a$ and $b$.}:
\begin{align}
	\delta_q S[q, t] &= \int_{s_i}^{s_f}\left(\PD{L}{q}\dot t-\frac{\dd}{\dd s}\PD{L}{(\dot q/\dot t)}\right)\cdot\delta q\,\dd s + \left.\PD{L}{(\dot q/\dot t)}\cdot\delta q\right\vert_{s_i}^{s_f}\notag\\
	\delta_t S[q,t] &= \int_{s_i}^{s_f} \left(\frac{\dd}{\dd s}\left[\PD{L}{(\dot q/\dot t)}\cdot\frac{\dot q}{\dot t}-L\right] + \PD{L}{t}\dot t\right)\delta t\, \dd s	-\left.\left(\PD{L}{(\dot q/\dot t)}\cdot\frac{\dot q}{\dot t}-L\right)\delta t\right\vert_{s_i}^{s_f}.
\end{align}
The usual assumption that the endpoints of the trajectory, i.e. $(q_i, t_i) := (q(s_i), t(s_i))$ and $(q_f, t_f) := (q(s_f), t(s_f))$, are held fixed under variation makes the boundary terms disappear. Simultaneously requiring that the action is stationary with respect to variations in $q$ and $t$ results in the equations of motion 
\begin{align}\label{eq:ContinuumEOM}
		\frac{\dd}{\dd s}\PD{L}{(\dot q/\dot t)}-\PD{L}{q}\dot t &=0\notag\\
		\frac{\dd}{\dd s}\left[\PD{L}{(\dot q/\dot t)}\cdot\frac{\dot q}{\dot t}-L\right] + \PD{L}{t}\dot t &= 0.
\end{align}
It is straightforward to show that the two equations of motion are not independent. More precisely, if $(q,t)$ satisfies the first one of these equations, then the second one is automatically satisfied and this is purely a consequence of the action's reparametrization invariance. This can be seen as follows: Let $\sigma:[s_i,s_f]\to I\subseteq\mathbb R$ be a $C^2$-map with $s\mapsto\sigma(s)$ and $\frac{\dd\sigma}{\dd s}\neq 0$ for all $ s\in[s_i,s_f]$\footnote{In concrete applications it might be more convenient to impose the stronger condition $\frac{\dd\sigma}{\dd s}>0$ or $\frac{\dd\sigma}{\dd s}<0$. The first condition preserves the orientation of the path while the second one reverses it, i.e. the trajectory $(q,t)$ is traversed from $(q_f,t_f)$ to $(q_i,t_i)$ which might be interpreted as a ``backward propagation in time''.}. Since the action functional~\eqref{eq:ActionFunctional} is clearly invariant under this reparametrization, it holds true that 
\begin{equation}
	\frac{\delta}{\delta\sigma}\int_{\sigma(s_i)}^{\sigma(s_f)} L\left(q(\sigma(s)), q'(\sigma(s))/ t'(\sigma(s)),t(\sigma(s))\right)\,  t'(\sigma(s))\,\dot\sigma(s)\,\dd s = 0,
\end{equation}
where the prime stands for a derivative with respect to $\sigma$ and the dot for a derivative with respect to $s$. An explicit computation of the variation leads to the identity
\begin{equation}
	\int_{\sigma(s_i)}^{\sigma(s_f)}\left[\PD{L}{q}\cdot q'+\PD{L}{(q'/t')}\cdot \frac{\dd}{\dd\sigma}\left(\frac{q'}{t'}\right)+\PD{L}{t}t'-\frac{\dd L}{\dd\sigma}\right]\, t'\,\dot\sigma\,\delta\sigma\,\dd s + L\, t'\,\delta\sigma\bigg\vert_{\sigma(s_i)}^{\sigma(s_f)} = 0.
\end{equation}
The integrand is obviously zero and the boundary term vanishes because the integration endpoints are fixed. However, assuming that $(q,t)$ is a solution of the first equation of motion~\eqref{eq:ContinuumEOM} in the parametrization $\sigma$ permits one to rewrite the integral as
\begin{equation}
	\int_{\sigma(s_i)}^{\sigma(s_f)}\left[\frac{\dd}{\dd\sigma}\left(\PD{L}{(q'/t')}\cdot\frac{q'}{t'}-L\right) + \PD{L}{t}t'\right]\, t'\, \dot\sigma\,\delta\sigma\,\dd s = 0,
\end{equation}
which is equivalent to $(q,t)$ satisfying the second equation of motion~\eqref{eq:ContinuumEOM}. Hence we have shown that the equations of motion are not independent due to the action's reparametrization invariance.

An immediate consequence of this fact is that the dynamical system is underdetermined: There are $n+1$ dynamical variables but only $n$ independent equations. It follows that a solution $(q,t)$ of~\eqref{eq:ContinuumEOM} is not uniquely determined by the boundary data, but rather one needs to provide a gauge fixing condition. For instance, one may impose the equation $\dot t(s) = h(s)$ for some function $h$ and subject this to the boundary condition $t(s_i) = t_i$. Then the first equation of~\eqref{eq:ContinuumEOM} and the boundary data can be used to determine $q$.\\
Evidently this procedure generates a particular solution $(q_h, t_h)$, which depends on the arbitrarily chosen gauge $h$ and which is therefore by no means unique. Any reparametrization\footnote{A reparametrization can be thought of as a ``change of gauge''.} of $(q_h, t_h)$ produces an equally valid solution to the equations of motion for the same boundary data. This means that $(q,t)$ are pure gauge variables and the equations~\eqref{eq:ContinuumEOM} really determine equivalence classes $([q],[t])$ of solutions. A generic solution $(q,t)$ of~\eqref{eq:ContinuumEOM} \textit{per se} is therefore physically meaningless. 

Notice however that \textit{all} representatives $(q,t)$ of a given equivalence class $([q],[t])$ determine \textit{the same} correlation between $q$ and $t$. That is to say that, at least locally, the arbitrary parameter $s$ can be eliminated and one obtains the gauge invariant function $q(t)$. This quantity is physically meaningful and we therefore refer to it as the physical solution.\\

To conclude this subsection we remark on an esthetic aspect:  It would be more appropriate to re-define the Lagrangian of the covariant theory to be
\begin{equation}
	\mathcal L(q(s), \dot q(s), t(s), \dot t(s)) := L(q(s), \dot q(s)/\dot t(s), t(s))\, \dot t(s).
\end{equation} 
This way, the equations of motion take on a more symmetrical form which manifestly treats $q$ and $t$ on the same footing:
\begin{align}\label{eq:CovEOM}
	\PD{\mathcal L}{q} - \frac{\dd}{\dd s}\PD{\mathcal L}{\dot q} &= 0\notag\\
	\PD{\mathcal L}{t} - \frac{\dd}{\dd s}\PD{\mathcal L}{\dot t} &= 0.
\end{align}
Since the Euler-Lagrange equations are covariant under point transformations of the dynamical variables, this justifies the use of the term \textit{covariant mechanics}.

\subsection{Symmetries and Noether's Theorem in the covariant Theory}\label{ssec:NoetherCont}

Consider a flow on the extended configuration space. That is, a one-parameter family of diffeomorphisms $\phi^\lambda: \mathcal C_\textsf{ext}\to\mathcal C_\textsf{ext}$ with $(q, t)\mapsto \phi^\lambda(q,t) =(\rchi^\lambda(q,t),\tau^\lambda(q,t))$ for all $\lambda\in\mathbb R$ and the group properties $\phi^0 = \text{id}$ and $\phi^\lambda\circ\phi^\mu = \phi^{\lambda+\mu}$ for all $\lambda,\mu\in\mathbb R$. Every flow possesses a generating vector field who's components are given by
\begin{equation}
	(u, v) := \left.\frac{\dd}{\dd\lambda}(\rchi^\lambda(q,t),\tau^\lambda(q,t))\right\vert_{\lambda=0}.
\end{equation}
Let $(q,t)$ be a trajectory in $\mathcal P$. Then a flow $\phi^\lambda$ maps this trajectory to $(q^\lambda(s), t^\lambda(s)) := \phi^\lambda(q(s),t(s))$. We call the flow $\phi^\lambda$ a continuous symmetry of the covariant Lagrangian $\mathcal L$ if
\begin{align}\label{eq:ContSymCond}
	\textcolor{white}{\Leftrightarrow} & \quad \mathcal L(q^\lambda, \dot q^\lambda, t^\lambda, \dot t^\lambda) - \frac{\dd}{\dd s}F(\lambda, q^\lambda, t^\lambda) = \mathcal L(q, \dot q, t, \dot t),
\end{align}
where $F$ is some function of $(\lambda, q,t)$ with the property $F\vert_{\lambda=0}=0$.
Notice that such a function can always be added to the covariant Lagrangian $\mathcal L$ without altering the equations of motion, provided it does not depend on the velocities $\dot q$ and $\dot t$. At this point it is convenient to introduce the conjugate momenta 
\begin{equation}\label{eq:Momenta}
	p := \PD{\mathcal L}{\dot q}\qquad\text{and}\qquad p_t := \PD{\mathcal L	}{\dot t},
\end{equation}
which allow the following concise formulation of Noether's theorem. \bigskip

\begin{center}
\parbox{0.92\textwidth}{\textbf{Covariant Version of Noether's Theorem:}\\ \textit{
	If $\phi^\lambda$ is a continuous symmetry of the covariant Lagrangian $\mathcal L$ and the trajectory $(q, t)$ satisfies the equations of motion~\eqref{eq:CovEOM}, then the quantity
	\begin{equation*}
		Q = p\cdot u + p_t\, v - \left.\frac{\dd}{\dd\lambda}F(\lambda, q^\lambda,t^\lambda)\right\vert_{\lambda=0}
	\end{equation*}
	is conserved along the whole trajectory $(q, t)$. That is to say, $\frac{\dd Q}{\dd s} = 0$ for all $s\in[s_i,s_f]$.
    }}
\end{center}\bigskip

The conventional form of Noether's theorem follows from the covariant version stated above when one chooses the gauge $t(s) = s$. Moreover, notice that $p$ and $p_t$ are gauge invariant quantities and therefore always correspond to the momentum and (the negative of) the energy of the physical solution. \bigskip

\begin{center}
\parbox{0.92\textwidth}{\textbf{Proof:}\\
Taking the derivative of $Q$ with respect to $s$ results in
\begin{align*}
	\frac{\dd Q}{\dd s} &= \dot p\cdot u + p\cdot\dot u	+ \dot p_t\, v + p_t\,\dot v - \left.\frac{\dd}{\dd s}\frac{\dd}{\dd\lambda}F(\lambda, q^\lambda, t^\lambda)\right\vert_{\lambda=0}\\
		&= \PD{\mathcal L}{q}\cdot u + p\cdot\dot u + \PD{\mathcal L	}{t}\,v + p_t\,\dot v - \left.\frac{\dd}{\dd\lambda}\frac{\dd}{\dd s}F(\lambda, q^\lambda, t^\lambda)\right\vert_{\lambda=0}\\
		&= \left.\frac{\dd}{\dd\lambda}\left(\mathcal L(q^\lambda, \dot q^\lambda, t^\lambda, \dot t^\lambda)-\frac{\dd}{\dd s}F(\lambda, q^\lambda, t^\lambda)\right)\right\vert_{\lambda=0} = 0.
\end{align*}
In the second line we made use of the equations of motion~\eqref{eq:CovEOM} expressed in terms of the conjugate momenta~\eqref{eq:Momenta} and in the last line we used the symmetry condition~\eqref{eq:ContSymCond}.$\hfill\blacksquare$
}
\end{center}\bigskip

\section{Discrete Lagrangian Formalism for Covariant Mechanics}\label{sec:Discrete}
There are many ways to discretize a continuum theory and some steps require the use of arbitrary prescriptions which we will now fix. This serves the purpose to have a well-defined framework within which the discrete version of Noether's theorem holds.

Consider the finite set $\{s_0, s_1,\dots, s_N\}\subset [s_i,s_f]$ supplemented by the conditions $s_0\equiv s_i$, $s_N\equiv s_f$, and $s_i < s_j$ for $i < j$. In particular, the elements of this set do not need to be equidistant. Based on this set, define the space of discretized trajectories as
\begin{equation}
	\mathcal{DP}^+ := \left\{(q,t):\{s_0,s_1,\dots, s_N\}\to\mathcal C_\textsf{ext} \, \vert\, (q,t)\in C^1(\mathbb R)\text{ and } t(s_{k+1})-t(s_k) > 0\right\}. 
\end{equation}
Notice that to define the space of trajectories $\mathcal P$ in the continuum it was sufficient to require $\dot t\neq 0$. That is because this condition together with the initial data $t_i<t_f$ implies forward propagation in $t$, i.e. $\dot t >0$. One may also choose the initial data such that $t_i>t_f$, which would imply $\dot t<0$, i.e a backward propagation in $t$.

To achieve the same in the discrete theory and, more importantly, to exclude certain spurious solutions of the discrete equations of motion, it is necessary to impose the condition $t(s_{k+1})-t(s_k)>0$. For discrete paths with backward propagation in time one would instead define a space $\mathcal{DP}^-$ where the last condition is replaced by $t(s_{k+1})-t(s_k)<0$. In what follows, we do not want to commit ourselves to a particular choice of ``time propagation'' and therefore simply write $\mathcal{DP}$, tacitly assuming that either $\mathcal{DP}^+$ or $\mathcal{DP}^-$ has been chosen. Moreover, we will identify discretized trajectories $(q,t)\in\mathcal{DP}$ with their image and write
\begin{align}
	(q,t) &\equiv \left(\{q(s_0), q(s_1),\dots, q(s_N)\}, \{t(s_0), t(s_1),\dots, t(s_N)\}\right)\notag\\
		&=: \left(\{q_0, q_1,\dots, q_N\}, \{t_0, t_1,\dots, t_N\}\right)\notag\\
		&=: \left(\{q_k\}, \{t_k\}\right).
\end{align}
The last line is a convenient shorthand notation which we will use from now on. We will sometimes visualize a discretized path as being a collection of \textit{vertices} in $(q,t)$ space labeled by an integer $k$, which are connected by \textit{line segments} or \textit{edges} $[k,k+1]$. To reflect whether a variable is associated with a vertex or a line segment, we write one index or two indices, respectively. A variable carrying two indices is for example the discretized tangent vector $(\dot q, \dot t)$. As derivatives with respect to $s$ are no longer defined, we choose to replace them by finite difference quotients which results in \begin{align}\label{eq:DifferenceQuotient}
	(\dot q, \dot t)\quad\longrightarrow\quad\left(\left\{\frac{\Delta q_{k-1,k}}{\Delta s_{k-1,k}}\right\}, \left\{\frac{\Delta t_{k-1,k}}{\Delta s_{k-1,k}}\right\}\right)\quad\text{with}\quad \Delta x_{k-1,k}:=x_k - x_{k-1}.
\end{align}
The integration measure $\dd s$ in~\eqref{eq:ActionFunctional} is not defined either and we choose to replace it by $\Delta s_{k-1,k}$. Taking into account the definitions given thus far, consider the discretized action functional $S_N:\mathcal{DP}\to\mathbb R$
\begin{align}\label{eq:DiscAction}
	S_N[\{q_k\}, \{t_k\}] :=& \sum_{k=1}^{N} L(q_{k-1}, \frac{\Delta q_{k-1,k }}{\Delta t_{k-1,k}}, t_{k-1})\Delta t_{k-1,k}= \sum_{k=1}^N L_{k-1,k}\Delta t_{k-1,k}.
\end{align}
Notice that the evolution parameter $s$ does not appear in the discretized action, as had already been observed in \cite{Rovelli2011} for the harmonic oscillator and emphasized in \cite{RovelliNewBook} for general Lagrangians. This can be taken as a hint that the discretization did not fully destroy reparametrization invariance. In fact, the continuum theory left some imprints of gauge invariance  which we will discuss in more detail in subsection~\ref{ssec:DELEQ}.

As expected, the discretized action functional depends on the whole discretized path $(\{q_k\},\{t_k\})$. From the action we can then easily identify the discretized Lagrangian as
\begin{equation}
	\mathcal L_{k-1,k}:= L_{k-1,k}\,\Delta t_{k-1,k} := L\left(q_{k-1},\frac{\Delta q_{k-1,k}}{\Delta t_{k-1,k}}, t_{k-1}\right)\,\Delta t_{k-1,k}.
\end{equation}
Notice that in the continuum the Lagrangian is a function of $s$ for a given choice of trajectory $(q,t)$. As such, it associates a number to every $s\in[s_i,s_f]$. In the discrete theory however, the Lagrangian associates a number not to a single value $s_k$ but to an interval $\Delta s_{k-1,k}$. We may therefore say that the discretized Lagrangian is associated to line segments $[k-1,k]$ of the discretization.

\subsection{Discrete Euler-Lagrange Equations}\label{ssec:DELEQ}
The first variation of the discretized action functional can be defined exactly as in the continuum and a straightforward computation leads to
\begin{align}\label{eq:Variations}
	\delta_q S_N[\{q_k\},\{t_k\}] &= \sum_{k=1}^{N-1}\left(\frac{\dd L_{k-1,k}}{\dd q_k}\Delta t_{k-1,k}+\frac{\dd L_{k,k+1}}{\dd q_k}\Delta t_{k,k+1}\right)\delta q_k \notag\\
	&\textcolor{white}{=\qquad}+\frac{\dd L_{0,1}}{\dd q_0}\Delta t_{0,1}\delta q_0 + \frac{\dd L_{N-1,N}}{\dd q_N}\Delta t_{N-1,N}\delta q_N\notag\\
	\delta_t S_N[\{q_k\},\{t_k\}] &= 	\sum_{k=1}^{N-1}\left(\frac{\dd L_{k-1,k}}{\dd t_k}\Delta t_{k-1,k}+L_{k-1,k}+\frac{\dd L_{k,k+1}}{\dd t_k}\Delta t_{k,k+1}-L_{k,k+1}\right)\delta t_k\notag\\
	&\textcolor{white}{=\qquad}+ \left(\frac{\dd L_{0,1}}{\dd t_0}\Delta t_{0,1}-L_{0,1}\right)\delta t_0 + \left(\frac{\dd L_{N-1,N}}{\dd t_N}\Delta t_{N-1,N}+L_{N-1,N}\right)\delta t_N.
\end{align}
Both variations contain two contributions which can be interpreted as the discrete analogues of the boundary terms $\PD{L}{\dot q}\delta q\vert_{s_i}^{s_f}$ and $-(\PD{L}{(\dot q /\dot t)}\frac{\dot q}{\dot t}-L)\delta t\vert_{s_i}^{s_f}$, respectively. These contributions disappear after imposing the boundary conditions $\delta q_0 = \delta q_N = 0$ and $\delta t_0=\delta t_N = 0$. However, these terms are interesting in their own right and we will come back to them in the next subsection.\\
What remains after imposing the boundary conditions and $\delta_q S_N = \delta_t S_N = 0$ are the discretized equations of motion
\begin{align}\label{eq:DELEQ_1}
	\frac{\dd L_{k-1,k}}{\dd q_k}\Delta t_{k-1,k}+\frac{\dd L_{k,k+1}}{\dd q_k}\Delta t_{k,k+1} &= 0\notag\\
	\frac{\dd L_{k-1,k}}{\dd t_k}\Delta t_{k-1,k}+L_{k-1,k}+\frac{\dd L_{k,k+1}}{\dd t_k}\Delta t_{k,k+1}-L_{k,k+1} &= 0.
\end{align}
At first sight these equations do not much resemble the continuum equations~\eqref{eq:ContinuumEOM}. This is merely due to our choice of variables. A form which more closely resembles the continuum equations can be obtained by rewriting the Lagrangian as
\begin{equation}\label{eq:RewrittenL}
	L_{k-1,k}(q_{k-1}, v_{k-1,k}, t_{k-1})\quad\text{with}\quad v_{k-1,k}:=\frac{\Delta q_{k-1,k}}{\Delta t_{k-1,k}}.
\end{equation}
With this Lagrangian, the discrete equations of motion~\eqref{eq:DELEQ_1} become
\begin{align}\label{eq:DiscreteEqs}
	\left(\PD{L_{k,k+1}}{v_{k,k+1}}-\PD{L_{k-1,k}}{v_{k-1,k}}\right)	 - \PD{L_{k,k+1}}{q_k}\Delta t_{k,k+1} &= 0\notag\\
	\left(\PD{L_{k,k+1}}{v_{k,k+1}}v_{k,k+1}-\PD{L_{k-1,k}}{v_{k-1,k}}v_{k-1,k}\right) - \left(L_{k,k+1}-L_{k-1,k}\right) + \PD{L_{k,k+1}}{t_k}\Delta t_{k,k+1} &= 0,
\end{align}
where we used $\PD{L_{k-1,k}}{q_k}=\PD{L_{k-1,k}}{t_k} = 0$, according to our discretization prescription (see \eqref{eq:RewrittenL}). This is a set of $(n+1)(N-1)$ independent equations supplemented by the $2(n+1)$ boundary conditions $(q_0, t_0)=(q_i, t_i)$ and $(q_N, t_N)=(q_f, t_f)$. This means that, for suitable boundary conditions, the equations~\eqref{eq:DiscreteEqs} completely determine the $(n+1)(N+1)$ variables $(\{q_k\},\{t_k\})$. This is in stark contrast to the continuum where the dynamical system is underdetermined.

Does this mean that reparametrization invariance is lost in the discrete theory? This depends on what one means by reparametrization invariance. As we have already seen in equation~\eqref{eq:DiscAction}, the discretized action is independent of the arbitrary evolution parameter $s_k$ and does, in this sense, not depend on any particular parametrization. As a consequence, the discrete equations of motion~\eqref{eq:DiscreteEqs} do not depend on the unphysical evolution parameter $s_k$ either. This means that a solution to these equations does not determine an arbitrarily parametrized path in $\mathcal C_\textsf{ext}$. Rather, it directly determines a correlation between the dynamical variables $q_k$ and $t_k$ and can therefore be regarded as the discretization of $q(t)$. Hence, the discrete theory directly determines (an approximation to) the physical solution.

The action's invariance with respect to reparametrization and the direct construction of the physical solution are key aspects which are retained in the discretized theory. In fact, the discrete theory does not require us to specify an arbitrary gauge fixing condition, the discrete equations of motion together with appropriate boundary conditions are enough to find the physical solution. Words like ``gauge'', ``equivalence classes'', and the like are absent in the vocabulary of the discrete theory. This means there is a whole lot of mathematical structure which is simply not needed to define and apply the discrete theory. In this sense, the discrete theory constitutes a minimal mathematical model of classical mechanics, lending further support to the ideas of \cite{Maudlin, Rovelli2019}.

\subsection{Conjugate Momenta and Boundary Terms}
In the continuum theory the definition of the conjugate momenta $p$ and $p_t$ involves derivatives with respect to $\dot q$ and $\dot t$, respectively. Since the dot refers to a total derivative with respect to the parameter $s$ it is not completely clear how to generalize this prescription to the discrete theory as there is no parameter $s_k$ and hence no obvious discrete analogue of $\PD{}{\dot q}$ and $\PD{}{\dot t}$. However, there is a simple alternative prescription.\\
Consider a trajectory $(q,t)$ which solves the continuum Euler-Lagrange equations with boundary conditions $(q_i, t_i)$ and $(q_f, t_f)$. It is well-known that for such a trajectory a variation of the Hamilton function with respect to the end points is, up to a sign, equal to the conjugate momentum $p$ and the energy $E=-p_t$. More precisely
\begin{align}
	\PD{S(q_i,t_i;q_f,t_f)}{q_i} &= -p_i & \PD{S(q_i,t_i;q_f,t_f)}{q_f} &= p_f\notag\\ 	
	\PD{S(q_i,t_i;q_f,t_f)}{t_i} &= E_i & \PD{S(q_i,t_i;q_f,t_f)}{t_f} &= -E_f,
\end{align}
where $(p_i, E_i)$, $(p_f,E_f)$ are the initial and final momenta and energies expressed as functions of the initial and final data of the trajectory $(q,t)$. By emulating these equations we can easily define initial and final momenta for a discretized path. To that end, we use the boundary terms appearing in the variations~\eqref{eq:Variations} to define
\begin{align}
	p^{i}_{0,1} &:= -\frac{\dd L_{0,1}}{\dd q_0}\Delta t_{0,1} & p^{f}_{N-1,N} &:= \frac{\dd L_{N-1,N}}{\dd q_N}\Delta t_{N-1,N}\notag\\
	E^{i}_{0,1} &:= \frac{\dd L_{0,1}}{\dd t_0}\Delta t_{0,1} - L_{0,1} & E^{f}_{N-1,N} &:= -\frac{\dd L_{N-1,N}}{\dd t_N}\Delta t_{N-1,N}-L_{N-1,N}.
\end{align}
 
\begin{figure}
	\centering
	\parbox{7.5cm}{
		\includegraphics[width=7.5cm]{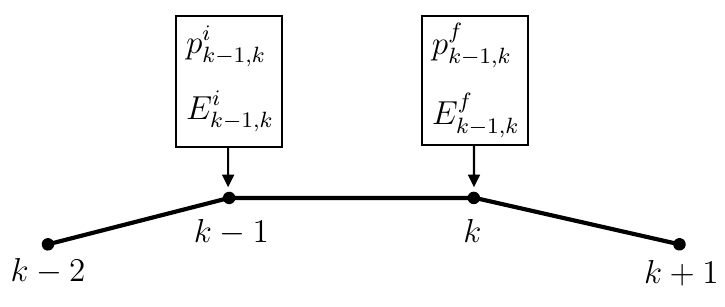}
		\caption{\textit{To every path segment belong two momenta and two energies.}}
		\label{fig:MomentaAndEnergies}}
		\qquad
	\begin{minipage}{7.5cm}
		\includegraphics[width=7.5cm]{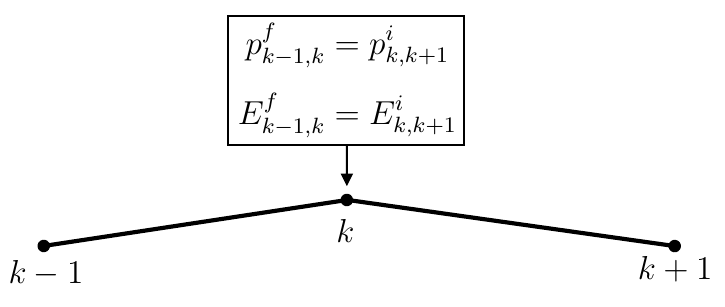}
		\caption{\textit{Energies and momenta between shared vertices agree due to the EOM.}}
		\label{fig:EOM}
	\end{minipage}
\end{figure}

Regarding every line segment $[k-1,k]$ of the discretized path $(\{q_k\},\{t_k\})$ as a path in its own right allows us to generalize the above definition such that we can assign energies and momenta not only to the initial and final vertex, but to all vertices $k$ along the discrete path:
\begin{align}\label{eq:EandP}
	p^{i}_{k-1,k} &:= -\frac{\dd L_{k-1,k}}{\dd q_{k-1}}\Delta t_{k-1,k} & p^{f}_{k-1,k} &:= \frac{\dd L_{k-1,k}}{\dd q_k}\Delta t_{k-1,k}\notag\\
	E^{i}_{k-1,k} &:= \frac{\dd L_{k-1,k}}{\dd t_{k-1}}\Delta t_{k-1,k} - L_{k-1,k} & E^{f}_{k-1,k} &:= -\frac{\dd L_{k-1,k}}{\dd t_k}\Delta t_{k-1,k}-L_{k-1,k}\quad \text{for } k\in\{1,\dots,N\}.
\end{align}
These prescriptions assign to every line segment two energies and two momenta. The line segment connecting the $(k-1)$-th and the $k$-th vertex, for instance, possesses an initial momentum $p^{i}_{k-1,k}$ and a final momentum $p^{f}_{k-1,k}$, see also Figure~\ref{fig:MomentaAndEnergies}. Since neighboring line segments share a vertex, these prescriptions also associate two momenta and two energies to each vertex. The $k$-th vertex, for example, possesses the momenta $p^{f}_{k-1,k}$ and $p^{i}_{k,k+1}$ and the energies $E^{f}_{k-1,k}$ and $E^{i}_{k,k+1}$. At first sight this seems to be inconsistent and indeed it might be inconsistent for unphysical discrete trajectories. However, there is no problem for paths $(\{q_k\},\{t_k\})$ which solve the equations~\eqref{eq:DELEQ_1}. In fact, in terms of the energies and momenta~\eqref{eq:EandP} the equations of motion~\eqref{eq:DELEQ_1} simply read
\begin{align}\label{eq:SimpleEOM}
	p^{f}_{k-1,k} &= p^{i}_{k,k+1}\notag\\
	E^{f}_{k-1,k} &= E^{i}_{k,k+1}.	
\end{align}
Hence, on physical trajectories the definitions~\eqref{eq:EandP} unambiguously associate one momentum and one energy to every vertex, see also Figure~\ref{fig:EOM}. Unphysical trajectories, on the other hand, are characterized by discontinuous behavior of the discrete energies and momenta at the vertices. The simple form \eqref{eq:SimpleEOM} of the equations of motion is also exactly what we need to prove a discrete version of Noether's theorem.

\subsection{Discrete Version of Noether's Theorem}
The same flow $\phi^\lambda:\mathcal C_\textsf{ext}\to\mathcal C_\textsf{ext}$ as in the continuum theory (see subsection~\ref{ssec:NoetherCont}) obeying the same group properties and possessing the same generating vector field can be applied in the discrete theory to transform the configuration space variables $(q, t)$ to $(\rchi^\lambda(q,t), \tau^\lambda(q,t))$. The only difference is how we denote the flow's action on trajectories:
\begin{equation}
	(\{\rchi^\lambda_k\},\{\tau^\lambda_k\}) := (\rchi^\lambda(\{q_k\},\{t_k\}), \tau^\lambda(\{q_k\},\{t_k\})) = \phi^\lambda(\{q_k\},\{t_k\})
\end{equation}
Under such a flow the discretized Lagrangian $L_{k-1,k}$ simply transforms as
\begin{equation}
	L^\lambda_{k-1,k} := L\left(\rchi^\lambda_{k-1}, \frac{\Delta\rchi^\lambda_{k-1,k}}{\Delta\tau^\lambda_{k-1,k}}, \tau^\lambda_{k-1}\right).
\end{equation}
We call $\phi^\lambda$ a continuous symmetry of $L_{k-1,k}\Delta t_{k-1,k}$ if, for all $\lambda\in\mathbb R$ and for all vertices $k$, the transformed Lagrangian satisfies the condition 
\begin{equation}\label{eq:SymCond}
	L^\lambda_{k-1,k}\Delta\tau^\lambda_{k-1,k} - \left(F_k - F_{k-1}\right) = L_{k-1,k}\Delta t_{k-1,k},
\end{equation}
where $F_k:=F(\lambda,\rchi^\lambda_k,\tau^\lambda_k)$ and $\left.F_k\right\vert_{\lambda=0}=0$\footnote{Notice that we can always add $(F_k-F_{k-1})$ to $L_{k-1,k}\Delta t_{k-1,k}$ without altering the discrete equations of motion as this only produces boundary terms which vanish due to the boundary conditions.}. An equivalent and more convenient formulation of the above condition is given by
\begin{align}\label{eq:EM_SymCond}
		&\quad \frac{\dd}{\dd\lambda}\left[L^\lambda_{k-1,k}\Delta\tau^\lambda_{k-1,k} - \left(F_k - F_{k-1} \right) \right] = 0\notag\\
		\overset{\text{at }\lambda=0}{\Longrightarrow} \quad &\quad p^{i}_{k-1,k}\cdot u_{k-1} - E^{i}_{k-1,k}\, v_{k-1} - \left.\frac{\dd}{\dd\lambda} F_{k-1}  \right\vert_{\lambda=0} = p^{f}_{k-1,k}\cdot u_k - E^{f}_{k-1,k}\, v_k - \left.\frac{\dd}{\dd\lambda} F_k  \right\vert_{\lambda=0}.
\end{align}
In the first line we silently assumed the boundary condition $[L^\lambda_{k-1,k}\Delta\tau^\lambda_{k-1,k} - \left(F_k - F_{k-1} \right) ]\vert_{\lambda=0} = L_{k-1,k}\Delta t_{k-1,k}$.
The second line in~\eqref{eq:EM_SymCond} renders the proof of Noether's theorem particularly simple which in the discrete covariant theory can be stated as follows.

\begin{center}
\parbox{0.92\textwidth}{\textbf{Discrete Version of Noether's Theorem:}\\ \textit{
	If $\phi^\lambda$ is a continuous symmetry of $L_{k-1,k}\Delta t_{k-1,k}$ and the trajectory $(\{q_k\},\{t_k\})$ satisfies the discrete equations of motion~\eqref{eq:SimpleEOM}, then the quantity
	\begin{equation*}
		Q_{k-1,k} = p_{k-1,k}^{f}\cdot u_k - E^{f}_{k-1,k}\, v_k - \left.\frac{\dd}{\dd\lambda}F_k\right\vert_{\lambda=0}
	\end{equation*}
	is conserved along the whole trajectory $(\{q_k\},\{t_k\})$. That is to say, $Q_{k-1,k}=Q_{k,k+1}$ for all $k\in\{1,\dots,N-1\}$.
    }}
\end{center}\bigskip

Notice that the definition of $Q_{k-1,k}$ given above seems to depend on the choice of vertex. Energy, momentum, the vector field $(u,v)$ and the function $F$ are all evaluated at $(q_k,t_k)$. Hence it seems that $Q_{k-1,k}$ is associated to the vertex $k$ and should, consistent with the notation used so far, carry a superscript $f$. However, $Q_{k-1,k}$ is really associated to the whole line segment $[k-1,k]$ and not to a particular vertex. To see this, rename the above quantity $Q^{f}_{k-1,k}$ and define  
\begin{align}
	Q^{i}_{k-1,k} := p_{k-1,k}^{i}\cdot u_{k-1} - E^{i}_{k-1,k}\, v_{k-1} - \left.\frac{\dd}{\dd\lambda}F_{k-1}\right\vert_{\lambda=0},
\end{align}
which is seemingly associated to the vertex $k-1$. It is easy to show, using solely the symmetry condition~\eqref{eq:EM_SymCond}, that $Q^{i}_{k-1,k}=Q^f_{k-1,k}$ and hence the definition of $Q_{k-1,k}$ does not depend on the choice of vertex. If on top of the symmetry condition also the equations of motion are satisfied, then $Q_{k-1,k}$ has the same value on all line segments of the discretized trajectory. This is the content of the theorem stated above which we will now prove.

\begin{center}
\parbox{0.92\textwidth}{\textbf{Proof:}\\
A straightforward computation yields
\begin{align*}
	Q_{k-1,k} &= 	p_{k-1,k}^f\cdot u_k - E_{k-1,k}^f\, v_k - \left.\frac{\dd}{\dd\lambda}F_k\right\vert_{\lambda=0}\\
		&= p_{k,k+1}^{i}\cdot u_k - E_{k,k+1}^{i}\, v_k - \left.\frac{\dd}{\dd\lambda}F_k\right\vert_{\lambda=0}\\
		&= p_{k,k+1}^f\cdot u_{k+1} - E_{k,k+1}^f\, v_{k+1} - \left.\frac{\dd}{\dd\lambda}F_{k+1}\right\vert_{\lambda=0} = Q_{k,k+1}.
\end{align*}
The first line is just the definition of $Q_{k-1,k}$. In the second line we used the equations of motion~\eqref{eq:SimpleEOM} and in the third line we made use of the symmetry condition~\eqref{eq:EM_SymCond}. It follows by induction that $Q_{k-1,k}=Q_{k,k+1}$ for all $k\in\{1,\dots,N-1\}$. $\hfill\blacksquare$
}
\end{center}\bigskip

It is interesting to remark that the theorem we just proved only holds in the discretized covariant theory. One can try to prove the same theorem starting from a discretization of the more familiar non-covariant continuum theory and  indeed it has been shown in \cite{Baez1994, Bahr2011} that it is possible to obtain a special case of Noether's theorem within this framework. However, the temporal variables are chosen such that $t_{k+1}-t_k = \epsilon$ for all $k$  and the diffeomorphisms are only allowed to act on the spatial variables $\{q_k\}$. Furthermore, any attempt to include flows acting on the  $\{t_k\}$ variables is bound to fail. The reason is that while it is possible to formulate an appropriate symmetry condition, it is not possible to relate energies between line segments sharing a vertex. In fact, the best one can do is to prove that the energy at the vertex $k$ computed from the left line segment is equal to the energy computed from the right line segment up to an unknown integration ``constant'' which depends on some of the $q_k$ and $t_k$ variables. Hence, in general there is a discontinuity between the energies and therefore energy cannot be a conserved quantity. 

This can ultimately be traced back to the fact that in the non-covariant approach the $\{t_k\}$ variables are chosen by hand and they are also subjected to the condition\footnote{Without this condition, conservation of momentum would also be violated.} $t_{k+1}-t_k = \epsilon$. It is easy to see that this choice does not affect the equations of motion \eqref{eq:DELEQ_1} resulting from the $q_k$ variation and therefore momentum conservation can be proven as it has been in \cite{Baez1994, Bahr2011} . However, this choice invalidates the equations of motion \eqref{eq:DELEQ_1} derived from the $t_k$ variation because they depend on a correct ``choice'' of $\Delta t_{k-1,k}$ and $\Delta t_{k,k+1}$. Hence, there is no hope for energy to be conserved and this also explains why the methods of \cite{Baez1994, Bahr2011} fail in the general case. 

A different situation presents itself in the discretized covariant theory. There, the spatial as well as the temporal variables are dynamical entities which are completely determined by the equations of motion. The dynamics is such that energies and momenta between shared vertices are related in the right way to make conserved quantities emerge. 

We illustrate this by applying the discrete Noether theorem to a few well-known examples.

\subsection{Examples: Galilean Transformations}
To illustrate the main result of this paper, we consider a covariant Lagrangian of the form
\begin{equation}\label{eq:modelLagrangian}
	\mathcal L_{k-1,k} = \frac{m}{2}\frac{(\Delta q_{k-1,k})^2}{\Delta t_{k-1,k}} - V\Delta t_{k-1,k},
\end{equation}
where $q_k\in\mathbb R^3$. The potential $V$ is not further specified, but it is assumed to be invariant (on a case by case basis) under the flows considered below and it may not only depend on $q_{k-1}$ but also on $v_{k-1,k}$, as it does for example for a particle in a magnetic field. We can then show that Galilean transformations are symmetries of this mechanical system and the conserved quantities are the discrete analogues of the quantities found in the continuum theory.\\

\underline{Translations in space:}\\
The flow $\phi^\lambda$ acts only on the position variables as $q_k\rightarrow\rchi^\lambda_k = q_k + \lambda a$ for some $a\in\mathbb R^3$. The generating vector field is given by $(u_k, v_k) = (a,0)$ and the corresponding conserved quantity is, as had to be expected, the conjugate momentum: $Q_{k-1,k} = p^{f}_{k-1,k}\cdot a = 	p^{i}_{k-1,k}\cdot a$.\\

\underline{Translations in time:}\\
Now the flow acts only on the time variables as $t_k\rightarrow \tau^\lambda_k = t_k+\lambda a$, for some $a\in\mathbb R$. The generating vector field reads $(u_k, v_k) = (0,a)$ and the conserved quantity is, of course, the energy: $Q_{k-1,k} = -E^f_{k-1,k}\, a = -E^{i}_{k-1,k}\, a$.\\

\underline{Rotations:}\\
Let $\phi^\lambda$ act on the spatial variables as $q_k\rightarrow \rchi^\lambda_k = R_{e}(\lambda) q_k$, where $R_{e}(\lambda)\in SO(3)$ is a rotation around the axis $e$ by an angle $\lambda$. The corresponding generating vector field is $(u_k, v_k) = (e\times q_k, 0)$, which yields the conserved quantity $Q_{k-1,k} = p^f_{k-1,k}\cdot (e\times q_k) = e\cdot(p^{f}_{k-1,k}\times q_k) = e\cdot(p^{i}_{k-1,k}\times q_{k-1})$. This is just the discrete analogue of angular momentum.\\    

\underline{Galilean boosts:}\\
A boost acts on the spatial variables as $q_k\rightarrow \rchi^\lambda_k = q_k + \lambda v t_k$, where $v\in\mathbb R^3$ is some constant velocity. This transformation does not leave the Lagrangian~\eqref{eq:modelLagrangian} invariant. Instead it leads to
\begin{align}
	L^\lambda_{k-1,k}\Delta t_{k-1,k} &= \frac{m}{2}\left[\frac{(\Delta q_{k-1,k})^2}{\Delta t_{k-1,k}} + 2\lambda v\cdot \Delta q_{k-1,k} + \lambda^2 v^2 \Delta t_{k-1,k}\right] - V\Delta t_{k-1,k}\notag\\
	&= L_{k-1,k}\Delta t_{k-1,k} + (F^\lambda_k-F^\lambda_{k-1}),
\end{align}
where $F^\lambda_k := \lambda m\, v\cdot q_k + \lambda^2 \frac{v^2 m}{2} t_k$. Hence, the symmetry condition~\eqref{eq:SymCond} is satisfied and we find the conserved quantity $Q_{k-1,k} = \left(p^{f}_{k-1,k}\, t_k - m\, q_k\right)\cdot v = \left(p^{i}_{k-1,k}\, t_{k-1}- m\, q_{k-1}\right)\cdot v$. Exactly the same as in the continuum. 

\section{Conclusion and Outlook}\label{sec:Conclusion}
Noether's theorem plays undeniably an important role in the discussion of mechanical systems. Here we showed that this powerful tool is also accessible for discretized mechanical systems, provided one starts from a covariant formulation of mechanics. In such a formulation, spatial and temporal variables $(q,t)$ are both dynamical and they are treated on an equal footing. Unlike in the continuum theory, however, the discretized equations of motion do not determine equivalence classes of solutions $([q],[t])$, but rather they explicitly determine all the variables $(\{q_k\},\{t_k\})$ and thereby directly determine an approximation to the physical solution $q(t)$. In the discrete covariant theory there is no need to talk about gauges or introduce arbitrary gauge fixing conditions. The absence of this mathematical structure, which in the continuum is ultimately only needed to deal with mathematical redundancy, can be taken as a sign that the discrete theory provides us with a minimal mathematical model of classical mechanics. This complements the ideas and observations of \cite{Maudlin, Rovelli2019} that the field theories of fundamental physics expressed in a discretized language need less structure for their definition than their continuum counterparts.

We have also seen that the dynamical determination of $(q_k, t_k)$ is also exactly what enables us to formulate a discrete version of Noether's theorem which is completely analogous to the well-known continuum version. More traditional discretizations of mechanical systems, such as the ones used in \cite{Baez1994, Bahr2011}, only allow for special cases of Noether's theorem. In particular, as we showed here, the special cases discussed in \cite{Baez1994, Bahr2011} exclude conservation of energy because fixing the $t_k$ variables by hand is tantamount to making the energy behave discontinuously on vertices of the discretized path. That a dynamical determination of the $t_k$ variables can lead to conservation of energy was first observed by \cite{Rovelli2011} in the study of the harmonic oscillator. The results presented here extend and clarify this observation in a broader context.

What we omitted here is to show how the discrete Noether theorem can be used to develop an algorithm to solve the discrete equations of motions in a way which renders transparent how the solutions depend on conserved quantities.  Such an algorithm would bring further theoretical applications of the discrete covariant theory into reach. However, before that,  important questions concerning the precise conditions under which solutions to the discrete equations of motion exist need to be answered.

\begin{acknowledgments}
	The author thanks Carlo Rovelli for helpful discussions and for reading an early draft of this paper. 
\end{acknowledgments}

\appendix
\section{From Covariant Mechanics to Maupertuis's Principle}\label{App_A}
Maupertuis's principle dates back to the 18$^{\text{th}}$ century and in its modern formulation it states that the configuration space trajectory $q(s)\subset\mathcal C$ of constant energy which connects $q_i$ to $q_f$ can be found as the stationary point of the abbreviated action functional
\begin{equation}
	S_0[q] := \int_{s_i}^{s_f} p\cdot \dd q.
\end{equation}
The parameter $s$ used to parametrize the trajectory $q$ is arbitrary and the abbreviated action is in fact reparametrization invariant.
It is usually assumed that the system also admits a Lagrangian description and that the Lagrangian has the form
\begin{equation}\label{eq:SpecialL}
	L = \frac{1}{2} g_{ab}(q)\dot q^{a} \dot q^{b} - V(q),
\end{equation}
where $g_{ab}$ is a Riemannian metric which only depends on $q$ and summation over repeated indices is implied. The variational principle can then be reformulated as
\begin{equation}
	\delta_q S_0[q] \equiv \delta_q \int_{s_i}^{s_f} \sqrt{2(E-V(q))}\, \dd l = 0,
\end{equation}
where $E$ denotes the constant energy of the trajectory and
\begin{equation}
	\dd l^2 = \frac12 g_{ab}(q)\, \dd q^{a}\dd q^{b}
\end{equation}
is the line element associated with the Riemannian manifold $(\mathcal C, g_{ab})$. This variational principle follows easily, in a slightly more general form, from the covariant language used in section \ref{sec:CovMech}. To see this, assume the Lagrangian is given by \eqref{eq:SpecialL}, with $\dot q$ replaced by $\dot q/\dot t$, and then take the variation of the covariant action functional \eqref{eq:ActionFunctional} with respect to $t$. As the Lagrangian is time-independent, the resulting equation of motion simply expresses conservation of energy and it can be integrated to
\begin{equation}
	\frac{1}{2}g_{ab}(q)\,\frac{\dot q^{a}}{\dot t}\frac{\dot q^{b}}{\dot t} + V(q) = E.
\end{equation}
Solving for $\dot t$ and choosing the positive square root (the choice of sign does not really matter due to Newtonian mechanic's invariance under time inversion) yields
\begin{equation}\label{eq:t_dot}
	\dot t = \sqrt{\frac{g_{ab}(q)\dot q^{a} \dot q^{b}}{2(E-V(q))}}.
\end{equation}
Substituting this expression back into the action functional results in
\begin{equation}\label{eq:M_Action}
	S[q] = \int_{s_i}^{s_f} \sqrt{2(E-V(q))}\,\dd l - (t_f-t_i)E.
\end{equation}
A variation of this action with respect to $q$ yields exactly the same equations of motion as Maupertuis's principle. However, while Maupertuis's principle only determines the shape of the trajectory, i.e. $q(s)$, in our approach we can also determine the correlation $q(t)$. To that end, vary the action \eqref{eq:M_Action} with respect to $E$ and set the variation to zero. This  results in
\begin{equation}
	t(s) = t_i + \int_{s_i}^{s}\frac{\dd l}{\sqrt{2(E-V(q))}},
\end{equation}
with $s\in[s_i, s_f]$. This is simply equation \eqref{eq:t_dot} in integral form and it allows us, in principle, to determine $q(t)$.

\newpage
\bibliographystyle{utcaps}
\bibliography{DiscreteNoether}

\end{document}